\documentclass[letterpaper]{emulateapj}

\usepackage{psfig}
\usepackage{rotating}

\newcommand{\Chandra}{{\em Chandra}}
\newcommand{\chandra}{\textit{Chandra}}

\newcommand{\xmm}{\textit{XMM-Newton}}

\newcommand{\nh}{\mbox{$N_H$}}

\newcommand{\kteff}{\mbox{$kT_{\rm eff}$}}
\newcommand{\rinfty}{\mbox{$R_{\infty}$}}

\newcommand{\nhtt}{\mbox{$N_{H,22}$}}
\newcommand{\chisq}{\mbox{$\chi^2_\nu$}}
\newcommand{\Chisq}[3]{$\chi^2_\nu$/dof (prob.) = {#1}/{#2} (#3)}

\newcommand{\xray}{\mbox{X-ray}}

\newcommand{\simlt}{\mathrel{\hbox{\rlap{\hbox{\lower4pt\hbox{$\sim$}}}\hbox{$<$}}}}
\newcommand{\simgt}{\mathrel{\hbox{\rlap{\hbox{\lower4pt\hbox{$\sim$}}}\hbox{$>$}}}}
\newcommand{\approxgt}{\mbox{$\,^{>}\hspace{-0.24cm}_{\sim}\,$}}
\newcommand{\approxlt}{\mbox{$\,^{<}\hspace{-0.24cm}_{\sim}\,$}}
\newcommand{\ee}[1]{\mbox{$10^{#1}$}}
\newcommand{\tee}[1]{\mbox{$\times 10^{#1}$}}
\newcommand{\ud}[2]{\mbox{$^{+ #1}_{- #2}$}}
\newcommand{\ppm}{\mbox{$\pm$}}

\newcommand{\unit}[1]{\mbox{$\rm\,#1$}}
\def\deg{\hbox{$^\circ$}}
\def\arcdeg{\hbox{$^\circ$}}
\def\arcmin{\hbox{$^\prime$}}
\def\arcsec{\hbox{$^{\prime\prime}$}}
\def\sec{\mbox{$\,{\rm sec}$}}

\newcommand{\km}{\hbox{$\,{\rm km}$}}

\newcommand{\keV}{\mbox{$\,{\rm keV}$}}
\newcommand{\eV}{\mbox{$\,{\rm eV}$}}
\newcommand{\ksec}{\mbox{$\,{\rm ks}$}}

\newcommand{\persec}{\mbox{$\,{\rm s^{-1}}$}}

\newcommand{\percmsq}{\mbox{$\,{\rm cm^{-2}}$}}
\newcommand{\percmcube}{\mbox{$\,{\rm cm^{-3}}$}}

\newcommand{\cgsflux}{\mbox{$\,{\rm erg\,\percmsq\,\persec}$}}
\newcommand{\cgslum}{\mbox{$\,{\rm erg\,\persec}$}}
\newcommand{\cgsdensity}{\mbox{$\,{\rm g\percmcube}$}}

\def\sourcefour{XMMU~171433$-$292747}       %
\def\sourcefive{XMMU~171421$-$292917}       %
\def\TwoMassfive{2MASS~17142095$-$2929163}  %
\def\sourceCeight{CXOU~J171432.93$-$292748.0}    %
\def\sourceCnine{CXOU~J171431.86$-$292745.5}     %
\def\sourceCseven{CXOU~J171420.88$-$292916.1}    %
\def\sourceCtwelve{CXOU~J171432.48$-$393735.3}   %

\begin{document}

\title{\Chandra Observation of Quiescent Low-Mass X-ray Binaries in the
Globular Cluster NGC~6304}

\author{Sebastien Guillot, Robert E. Rutledge} \affil{Department of
Physics, McGill University,\\ 3600 rue University, Montreal, QC,
Canada, H2X-3R4} \email{guillots@physics.mcgill.ca,
rutledge@physics.mcgill.ca} \author{Edward F. Brown} \affil{Department
of Physics and Astronomy, Michigan State University, 3250 Biomedical
Physical Science Building, East Lansing, MI 48824-2320, USA}
\author{George G. Pavlov} \affil{Pennsylvania State University, 512
Davey Lab, University Park, PA 16802, USA} \author{Vyacheslav
E. Zavlin} \affil{Space Science Laboratory, Universities Space
Research Association, NASA MSFC VP62, Huntsville, AL 35805, USA}

\slugcomment{Draft submitted to ApJL on \today}
\shorttitle{Chandra observations of qLMXBs in NGC 6304}

\begin{abstract}
This paper presents the analysis of candidate quiescent low-mass \xray\ binaries 
(qLMXBs) observed during
a short \Chandra/ACIS observation of the globular cluster (GC)
NGC~6304.  Two out of the three candidate qLMXBs of this cluster,
\sourcefour\ and \sourcefive, lie within the field of view.  This
permits comparison with the discovery observation of these sources.
The one in the GC core -- \sourcefour\ -- is spatially resolved into
two separate X-ray sources, one of which is consistent with a pure
H-atmosphere qLMXB, and the other is an X-ray power-law spectrum
source.  These two spectral components separately account for those
observed from \sourcefour\ in its discovery observation.  We find that
the observed flux and spectral parameters of the H-atmosphere spectral
components are consistent with the previous observation, as expected
from a qLMXB powered by deep crustal heating.  \sourcefive\ also has
neutron star atmosphere spectral parameters consistent with those in
the \xmm\ observation and the observed flux has decreased by a factor
$0.54\ud{0.30}{0.24}$.
\end{abstract}


\maketitle


\section{Introduction}
The faint emission of transiently accreting low-mass \xray\ binaries
in quiescence (qLMXBs) found in globular clusters (GCs) is often used
to measure the radii of neutron stars (NSs).  Such measurements, free
of large uncertainties caused by unknown distances and atmospheric
composition, can provide useful constraints on the nuclear dense
matter equation of state (EoS) relating pressure and density when
matter has a density above $2.35\tee{14}\cgsdensity$, such as can
occur in atomic nuclei, and in the interiors of NSs
\citep{lattimer04}.  Obtaining precise constraints on the dense matter
EoS is the observational motivation for NS radii measurements,
requiring at least a $\sim5\%$ accuracy to be useful
\citep{lattimer04}.

The expected overabundance of LMXBs in GCs \citep{hut92} has
motivated observations with the current generation of \xray\
telescopes.  It was then empirically shown that the NS binaries
population in GC depends on the interaction rate of the cluster
\citep{gendre03b,heinke03b,pooley03}.  Several qLMXBs were discovered
close to the cores of the GCs \citep[for example,][]{rutledge02b}.  In
most cases, they were spectrally identified based on their spectra
consistent with a NS atmosphere at the distance of their host cluster.
The known distances to the clusters and the known values of the
absorption led to precise NS radii measurements \citep[$\sim 5-20\%$
uncertainty, ][]{rutledge02b, gendre03b, heinke03c}.  However, only a
handful of confirmed GC qLMXBs are known, and finding more of those
objects necessitates careful \xray\ observations of GCs.  Finally, the
X-ray luminosity qLMXBs is not expected to show variability on
$\sim$years timescales \citep{brown98,ushomirsky01}, but long-term
flux variation have been observed before \citep{rutledge02a}.  A more
complete introduction about qLMXBs and an exhaustive list of GC
qLMXBs can be found elsewhere \citep[G09 hereafter]{guillot09a}.

Regarding the observations of GCs, the two \xray\ telescopes \xmm\ and
\Chandra\ are complementary.  On the one hand, the typical luminosity
of qLMXBs ($L_{\rm X}\sim10^{32}-10^{33} \cgslum$) and the typical
distances to GCs require the effective area of \xmm\ to collect high
signal-to-noise ratio (S/N) data for spectral analyses, within modest
integration times.  On the other hand, \Chandra's angular resolution
permits spatial resolution of adjacent sources in cores of GC, which
\xmm\ is not able to differentiate otherwise.

We present here a comparative analysis of an archived \Chandra\
observation of NGC~6304, with the results of the \xmm\ data reported
earlier this year.  The sole focus of this paper is the analysis of
the previously reported qLMXBs (G09).  Two out of the three identified
candidate qLMXBs lie within the field of view of this observation:
\sourcefour\ and \sourcefive.  In Section~\ref{sec:red}, we describe
the observation, data reduction, source detection, and approach for the
spectral analysis of these two sources.  Section~\ref{sec:comp}
contains the actual analysis, with positional and spectral
comparisons.  Section~\ref{sec:discuss} includes a discussion and a
short conclusion.

\section{Data Reduction and Analysis}
\label{sec:red}

\subsection{Observation and Data Reduction}
\label{sec:obs}

We analyzed here an archived \Chandra\ observation of NGC~6304
(Table~\ref{tab:Obs}).  The CIAO V4.1.1 \citep{fruscione06} package is
used for the source detection and data analysis. The pre-processed
event file (level 2) is analyzed including events in the 0.5--8.0\keV\
range.  The data are checked for flares and none are found.  The low
luminosity of \xray\ sources in this GC allows us to safely neglect
pile-up; the brightest source, accounting for $\sim80$ counts in
total, corresponds to a pile-up fraction of less than
1\%\footnote{\Chandra\ Observatory Proposer Guide, chap. 6, v11.0,
January 2009}.

\begin{deluxetable}{rr}
  \tablecaption{\label{tab:Obs} \Chandra\ Observation of NGC~6304}
  \tablewidth{0pt}
  \tabletypesize{\scriptsize}    
  \tablecolumns{2}
  \tablehead{
    \multicolumn{2}{c}{Parameters of the Observation}
  } 
  \startdata 
  Obs. ID & 8952 \\   
  Starting time &  2008 Jan. 28 18:10:38 (TT)\\ 
  Exposure time & 5262\sec\\
  Detector & ACIS-S3 (BI)\\ 
  Frame rate & 3.1410\sec\\ 
  \enddata 
\end{deluxetable}

The source detection is performed with the {\tt wavdetect} algorithm,
treating each ACIS chip separately, using the following parameters: an
exposure threshold {\it expthresh = 0.1} and the wavelet scales {\it
scales = ``1.0 2.0 4.0 8.0''}.  Six sources (with significance
$\sigma>3$) are detected on the ACIS-S3, plus a low-significance
($2.3\sigma$) source, located in the core of NGC~6304.  Eight other
sources ($\sigma>3$) are found on other operative chips (two on
ACIS-I3, two on the edges of ACIS-S1, three on ACIS-S2 and one on
ACIS-S4).  Also, we perform the detection over two narrower energy
ranges: 0.5--2\keV\ and 2--8\keV.  In the 0.5-2\keV\ data, we detect
in the GC core no additional source with significance greater than
$3\sigma$.  In the 2--8\keV\ data, only source C09 is detected in the
core with $>3\sigma$ significance.

For comparison, the detection is also performed on the full
0.5--8\keV\ using the exposure map (created with {\tt mkexpmap}
following the analysis thread \emph{``Single Chip ACIS Exposure
  Map''}).  Similar detections are obtained, except for a few minor
differences: the two sources located near the edges of ACIS-S1 are not
detected, suggesting that they are false detections.  The
low-significance source ($2.3\sigma$), detected in the core of the GC,
on ACIS-S3, is not found with this detection method.  Other than these
differences, the two detection runs found the same sources.  The
statistical positional uncertainties obtained with the previous
detection run (without the exposure map) are smaller than that
obtained when the exposure map is used. In consequence, the results
with the smallest statistical uncertainties (detection without the
exposure map) are presented and used.  Table~\ref{tab:pos} shows the
results of the source detection for the sources of interest of this
paper.  The uncertainty in the source positions is the quadratic sum
of the statistical uncertainty ($\sim0.1\arcsec$) and the systematic
positional uncertainty of $\sim0.6\arcsec$ for \Chandra\
\footnote{From the \Chandra\ Calibration web page available at
  {http://cxc.arvard.edu/cal/}}.

\subsection{Source Extraction and Spectral Analysis}
\label{sec:spectra}

The focus of this paper is the confirmation of the detection of the
qLMXBs in the field of the GC NGC~6304.  Therefore, the analysis
presented afterward pertain solely to the two discovered qLMXBs
(G09) that lie in the field of view of this \Chandra\ observation.

The script {\tt psextract}, together with the calibration files from
CALDB v4.1 \citep[containing the latest effective area maps, quantum
efficiency maps and gain maps]{graessle07}, is used to extract the counts
of the \xray\ sources, in the energy range 0.5--8.0\keV.  In all
cases, the extraction region is chosen so that more than 90\% of the
energy is included (encircled count fraction $ECF>90\%$ \footnote{The
extraction radii are determined using the webtool available at:
http://cxc.harvard.edu/cgi-bin/build\_viewer.cgi?psf}).  For sources
at small off-axis angle, i.e., sources in the core of the cluster, the
circular extraction region of radius 2.5\arcsec\ around a source
comprises more than 95\% of the source energy\footnote{Chandra
Observatory Proposer Guide, chap. 6, v11.0, January 2009}.

Since this observation was performed with the ACIS-S instrument and a
focal plane temperature of -120\unit{\deg C}, the response matrices
files (RMF) have to be recalculated, according to the recommendations
of the CIAO Science Thread \emph{``Creating ACIS RMFs with
mkacisrmf''}. It is also crucial to recalculate the ancillary response
file (ARF) using the new RMFs in order to match the energy grids
between the RMF and ARF files.

Due to the low count statistic in this observation, the spectra are
left unbinned, and the Cash-statistic \citep{cash79} permits to find
the best-fit parameters, assuming the fitted model is correct, using
the software XSPEC v12.3 \citep{arnaud96}.  The NS atmosphere model
used here is a tabulated model similar to the XSPEC model {\tt nsa}
\citep{zavlin96}, but without the hard limits on the projected radius
imposed by the {\tt nsa} model.  The results of the spectral fits are
listed in Table~\ref{tab:res}.

\section{Comparison with the \xmm\ Observation}
\label{sec:comp}

\subsection{In the core of NGC~6304 - \\ \sourcefour}
\label{sec:core}

This object was the sole source discovered with \xmm\ in the core of
NGC~6304 (G09), and was classified as a qLMXB based on its X-ray
spectrum.  It was composed of two components: an H-atmosphere neutron
star spectrum at the distance of NGC~6304, and a hard power-law
component which dominates the spectrum at high energies.  A marginal
positional offset between the thermal source (candidate qLMXB)
detected on a soft image ($<1.5\keV$) and a harder source (power-law
component) detected on a hard image ($>1.5\keV$) of the \xmm\ data
tentatively suggested that the two spectral components are due to
distinct point sources (G09).

It is shown in the following subsections that the \xmm\ source
\sourcefour\ is resolved, spatially and spectrally, into two sources
detected with \chandra.

\subsubsection{Positional analysis}

\begin{figure}
  \centerline{~\psfig{file=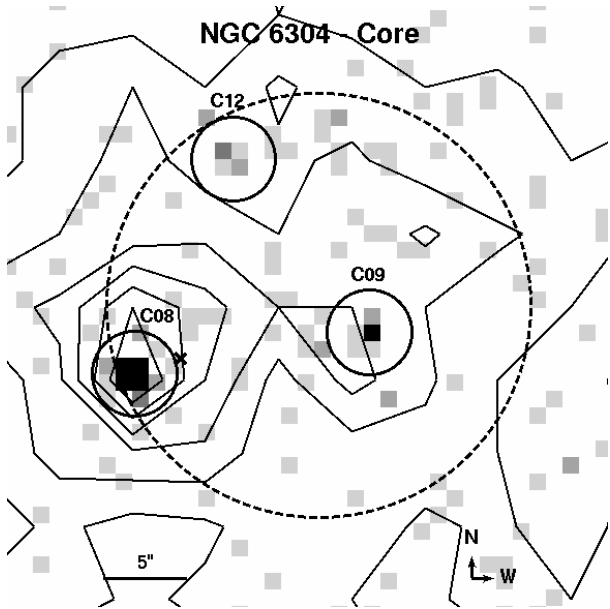,width=8cm}~}
  \bigskip
  \caption[]{This \Chandra/ACIS-S3 image of the core of the globular
    cluster NGC~6304 was created from the level-2 pre-processed file
    in which the events were binned by a factor $\times 2$.  The
    dashed line represents the core radius, $r_{\rm c}=0.21\arcmin$
    \citep{harris96}.  The cross shows the position of the previously
    detected source \sourcefour, and the contour lines represent the
    overlapped event distribution of the \xmm\ data.  The {\em XMM} contour
    have been shifted by 2.6\arcsec\ to the west and 1.1\arcsec\ to
    the north.  The total shift of 2.8\arcsec\ is consistent with the
    relative pointing accuracy of 1.6\arcsec ($1\sigma$) between the
    two observations (see Section~\ref{sec:core}). Finally, the three solid
    circles are the 2.5\arcsec\ extraction radii of the \Chandra/ACIS
    sources.  C12 was detected with a low significance: $2.3\sigma$.
    \label{fig:core}}
\end{figure}

In Figure~\ref{fig:core}, two sources, \sourceCeight\ and \sourceCnine,
are detected ($\sigma>3$) within the core radius of NGC~6304,
separated by an angular distance of $\sim13\arcsec$.  The brightest
one -- \sourceCeight, accounting for 74 counts (including about two
background counts) -- is positionally consistent with the candidate
qLMXB \sourcefour, within $2\sigma$.  The relative offset between the
\xmm\ and \Chandra\ positions of the source is
2.9\arcsec\ppm1.6\arcsec; the error on the \xmm\ astrometry
(statistical and systematic errors) accounts for 1.5\arcsec\,
($1\sigma$).  Since \xmm\ cannot resolve the two sources in the core,
the centroid position obtained with {\tt wavdetect} is in between the
two sources detected with \Chandra/ACIS (Figure~\ref{fig:core}).

\subsubsection{Spectral characterization and flux calculation of
  \sourceCeight\ and \sourceCnine}

As mentioned above, the extracted events are left unbinned and the
spectrum is fitted with the assumed model, including a 3\% systematic
uncertainty, and the confidence regions are derived using the
Cash-statistic \citep{cash79}.  The galactic absorption (hydrogen
column density, \nh) is kept fixed at the value in the direction of
NGC~6304: $\nh=0.266\tee{22} \unit{atoms\percmsq}$ ($\nhtt=0.266$
afterward).  The derived best-fit parameters and confidence regions
(Table~\ref{tab:res}) are compared to the previously reported values
(G09).

The source \sourceCeight\ is spectrally soft with all events having
energies below 2.5\keV.  It is assumed to correspond to the qLMXB
\sourcefour, and therefore a NS hydrogen atmosphere model is used for
the spectral fitting.

For this \xray\ source -- the only one in this observation -- the
number of counts allows for spectral binning ($\sim15$ counts per
bins), while having an acceptably approximate Gaussian uncertainty in
each bin.  The chosen background is a circular region of 1\arcmin\
around the core, excluding 5\arcsec\ ($ECF=99\%$) around each detected
source.  A 5-bin spectra is obtained in the 0.5-8.0\keV\ range.  The
model used is the tabulated model of NS hydrogen atmosphere
\citep{zavlin96}, together with fixed $\nhtt=0.266$.  The best-fit
parameters of this \chisq-fit are: $\kteff=128\ud{40}{30}\eV$ and
$\rinfty=7.3\ud{10.5}{3.7}\km$.  Although the counting statistic is
poor, the best-fit values are in agreement with the previous results,
and the model fitted is statistically acceptable
(\Chisq{0.12}{3}{0.95}).  For completeness, the spectrum is fitted with
an absorbed power law, leading to a best-fit photon index
$\alpha=3.5\ud{0.5}{0.4}$ with \Chisq{1.3}{3}{0.27}, which indicates a
soft source, as expected for a typical qLMXB.  This provides further
support to the argument that \sourceCeight\ is the qLMXB detect in
\sourcefour.  In Table~\ref{tab:res}, we report best-fit values
determined using not the $\chi^2$ statistic with binned data, but the
Cash-statistic using unbinned data, which typically produces smaller
error regions.  The results of Monte Carlo ``goodness-of-fit''
simulation are also given in this table.

\begin{deluxetable*}{clccccc}[b]
  \tablecaption{\label{tab:pos} Sources of interest}
  \tablewidth{0pt}
  \tabletypesize{\scriptsize}
  \tablehead{
    \colhead{Object Name} & \colhead{ID} & 
    \colhead{$\delta_{\rm R.A.}\backslash \delta_{\rm decl.}$} & 
    \colhead{$\sigma$} & \colhead{Counts} & 
    \colhead{{\em XMM} Name} & \colhead{{\em XMM} ID}
  }
  \startdata 
  \sourceCeight & C08 (core) & 0.08$\backslash$0.08 & 24.2 & 72 &\sourcefour & \#04 \\
  \sourceCnine  & C09 (core) & 0.17$\backslash$0.12 & 7.05 & 16 &\sourcefour & \#04 \\
  \sourceCseven & C07        & 0.19$\backslash$0.15 & 6.4  & 16 &\sourcefive & \#05 \\
  \sourceCtwelve& C12 (core) & 0.27$\backslash$0.17 & 2.3  & 5  &\sourcefour & \#04 \\
  \enddata 
  \tablecomments{$\delta_{\rm R.A.}$ and $\delta_{\rm decl.}$ are the
  statistical uncertainties from the source detection, in seconds of
  arc.  Systematic position uncertainties consist of an additional
  error of 0.6\arcsec\ (90\% confidence, see Section~\ref{sec:obs}).  The
  column ``counts'' reports the background subtracted number of counts
  for each source, in the range 0.5--8\keV.  Finally, $\sigma$
  represents the significance of the source detection.}
\end{deluxetable*}

The best-fit H-atmosphere parameters for \sourceCeight\ are consistent
with typical values for quiescent NSs (G09, Table 4 for exhaustive
listing of known GC qLMXBs), and with the best-fit values obtained
with \xmm\ (G09).  The unabsorbed flux (0.5--10\keV) is also
consistent with the flux of the thermal component of the \xmm\
observation (as will be demonstrated statistically below), estimated
to be 51\% of the total flux (G09), the other 49\% being the
contribution of the power-law component.

The second source in the core, \sourceCnine\ (18 counts including
about two background counts), appears spectrally harder, as it has
only seven counts below 2\keV, two counts in the 2.0--3.0\keV\ energy
range, and the rest (nine events) above 3\keV.  The spectrum is fitted with
an absorbed power law, to compare it to the power-law component fitted
in the spectrum of \sourcefour.  The best-fit photon index is
consistent with the hard power-law component for the source
\sourcefour .  For completeness, the spectrum is fitted with a NS
atmosphere model, for which the best-fit projected radius is
inconsistent with the typical radii of NSs.  The goodness of this fit
(99.9\% of Monte Carlo simulations from the NS atmosphere model give
better statistics than the best fit) suggests that the spectrum is not
that of a NS H-atmosphere model.  These results provide further
support that this source is not the candidate qLMXB, but another
source of unknown classification.

As a last check of statistical consistency, a simultaneous fit is
performed using the {\em XMM}/pn, {\em XMM}/MOS1, {\em XMM}/MOS2 for
\sourcefour\ and \Chandra/ACIS spectra for both \sourceCeight\ and
\sourceCnine.  While fitting, the temperature, the radius and the
photon index parameters of each individual data set are kept tied
together and \nh\ is kept fixed at the value cited above.  The results
are also shown in Table~\ref{tab:res}, as a simultaneous fit for
\sourcefour.  Again, the fit is statistically acceptable,
\Chisq{0.76}{42}{0.86}, and the obtained best-fit parameters are in
agreement with typical values for accreting quiescent NS.

To characterize the variation in flux for the qLMXB, we performed a
simultaneous spectral fitting using the EPIC/pn data alone for
\sourcefour\ and with the ACIS-S data for \sourceCeight\ and
\sourceCnine; a multiplicative factor for the spectral normalization
is used, fixed at 1 for the \xmm/pn spectrum and left as a free
parameter for the \Chandra/ACIS spectrum.  The best-fit factor is
$0.80\ud{0.19}{0.16}$(90\% confidence), which is marginally consistent
with the fluxes being the same.

We therefore conclude that \sourcefour, observed previously in the
core of NGC~6304 (G09), is a composite of \sourceCeight\ and
\sourceCnine, which were not distinguishable at the resolution of
\xmm, but which are spatially resolved at the resolution of \chandra.

\subsection{Outside the core of NGC~6304 - \\ \sourcefive}
\label{sec:rest}

The low signal-to-noise candidate qLMXB \sourcefive\ reported in the
\xmm\ analysis (G09) is also detected on the \Chandra/ACIS observation
as \sourceCseven.  An offset of 1.15\arcsec\ppm1.6\arcsec\ is measured
between the \chandra/ACIS and the \xmm\ observations, consistent
within $1\sigma$.  The \chandra\ source position is located
0.9\arcsec\ppm 0.6\arcsec\ from the possible Two-Micron All Sky Survey
(2MASS) counterpart reported \TwoMassfive\ (G09).  This offset is
consistent ($1.4\sigma$) with the two sources (\sourceCseven\ and
\TwoMassfive ) being associated where the uncertainty is due to
\chandra's systematic and statistical uncertainties; the error on the
2MASS position is assumed to be negligible.  Also, the probability
that another source as bright or brighter lies as close or closer to
the \xray\ positions is 0.34\%, providing further support to the
association, with 99.66\% confidence.

This \xray\ source, \sourceCseven, is located $\sim1.5\arcmin$
off-axis and requires a 3\arcsec -radius extraction region.  The
source has 19 counts (including about three background counts) and is
also fitted, using Cash-statistic, with a NS atmosphere model, keeping
$\nhtt=0.266$ fixed.  The best-fit parameters are consistent with the
previously published values (Table~\ref{tab:res}).

A simultaneous fitting is also performed for this candidate qLMXB,
using the \xmm\ and \chandra\ spectra.  The \chandra/ACIS spectrum has
only one bin, containing 19 counts.  The fit is statistically
acceptable (using $\chi^2$ statistics) and the best-fit parameters are
consistent with those published elsewhere (G09).

Using the same method as described in Section~\ref{sec:core}, the
simultaneous fitting with a multiplication factor suggests that the
flux has changed between the \xmm\ observation and the \Chandra/ACIS
observation.  Indeed, as the factor mentioned is fixed to 1 for the pn
spectrum, the best-fit factor is $0.54\ud{0.30}{0.24}$(90\%) for the
ACIS spectrum.  Higher signal-to-noise data will permit confirmation
of the apparent variability in the flux of this candidate qLMXB.

\begin{deluxetable*}{ccccccc}
  \tablecaption{\label{tab:res} Spectral Results}
  \tablewidth{0cm}
  \tabletypesize{\scriptsize}
  \tablecolumns{7}
  \tablehead{
    \colhead{ID} & \colhead{\nhtt} & \colhead{\kteff} & 
    \colhead{\rinfty} & \colhead{$\alpha$} & \colhead{$F_{-13,X}$} &
    \colhead{\chisq/d.o.f. (prob.)}\\
    \colhead{} & \colhead{} & \colhead{(eV)} & \colhead{(km)} & 
    \colhead{} & \colhead{} & \colhead{}  
  }
  \startdata 
  \sourceCeight & (0.266) & 127\ud{34}{27} & 7.5\ud{8.3}{3.7}  & \nodata          & 1.14\ud{0.24}{0.21} & Goodness: 35.0\%\\
      ''        & (0.266) & \nodata        & \nodata           & 3.5\ud{0.4}{0.5} & \nodata             & Goodness:  9.3\%\\
  \sourceCnine  & (0.266) & \nodata        & \nodata           & 0.8\ud{0.7}{0.7} & 0.72\ud{0.33}{0.67} & Goodness: 76.2\%\\
      ''        & (0.266) & $\sim500\eV$   & $<1\km$           & \nodata          & \nodata             & Goodness: 99.9\%\\
  \sourceCseven & (0.266) &  89\ud{49}{46} &  9.3\ud{410}{4.6} & \nodata          & 0.32\ud{0.13}{0.11} & Goodness: 77.2\%\\
  \sourceCtwelve\tablenotemark{a}& (0.266) & \nodata & \nodata & (2)              & 0.15                & webPIMMS\\
  \cutinhead{Spectral results from G09}
  \sourcefour   & (0.266) & 122\ud{31}{45} & 11.6\ud{6.3}{4.6} & 1.2\ud{0.7}{0.8} & 2.3     & 0.85/42 (0.75) \\
  \sourcefive   & (0.266) &  70\ud{28}{20} & 23\ud{69}{10}     & \nodata          & 0.6     & 1.09/16 (0.36) \\ 
  \cutinhead{Simultaneous fitting with XMM/EPIC and \Chandra/ACIS  spectra}
  \sourcefour   & (0.266) & 123\ud{23}{25} & 7.9\ud{6.4}{2.2}  & 0.3\ud{0.8}{1.4} & 
                2.0--3.2 \tablenotemark{b} & 0.76/42 (0.86) \\
  \sourcefive   & (0.266) &  65\ud{23}{16} & 30\ud{66}{15}     & \nodata          & 
                0.36--0.66 \tablenotemark{b} & 1.11/22 (0.33) \\
\enddata 

  \tablecomments{This table presents the spectral results of the
    applied model: NS atmosphere or power law, for which the
    parameters are quoted: \kteff\ and \rinfty\ or photon index
    $\alpha$, respectively.  The unabsorbed flux $F_{-13,X}$ is
    expressed in units of $\ee{-13}\cgsflux$ (0.5--10\keV), and the
    errors are estimated using the XSPEC convolution model {\tt
    cflux}.  In all cases, the absorption, \nh, is kept fixed at the
    value $\nh=0.266\tee{22} \unit{atoms\percmsq}$.  "Goodness"
    indicates that the fit was performed with Cash-statistic and the
    results of Monte Carlo simulations of the goodness of fit are
    provided.  A percentage close to 50\% suggests a good fit while
    extreme values indicate poor fits.}  
  \tablenotetext{a}{This low-significance source (C12) is given for
   information purposes due to its proximity to the candidate qLMXB in
   the core.  The flux is estimated using webPIMMS, assuming a
   power law of photon index $\alpha=2$.}  
  \tablenotetext{b}{Each component of the simultaneous fitting has its
   own estimated model flux. The range is given here.}
\end{deluxetable*}

\section{Discussion and Conclusions}
\label{sec:discuss}
We have performed the spectral analysis of two candidate qLMXBs in
NGC~6304 -- \sourcefour\ and \sourcefive\ -- detected on a 5.2\ksec\
\Chandra/ACIS observation.  The third reported candidate qLMXB was
outside of the field of view of telescope.  As suggested previously,
the candidate \sourcefour\ is resolved into two \xray\ sources:
\sourceCeight, a bright ($L_{\rm X}=0.5\tee{33}\cgslum$) thermal
source and \sourceCnine, a second fainter harder source of unknown
classification.  The spectrum of the \sourceCeight\ was fitted with a NS
atmosphere model at the distance of NGC~6304 and the best-fit
parameters are consistent with the parameters obtained from the fit of
the \xmm\ data (G09).  The spectrum of \sourceCnine\ was fitted with a
simple absorbed power-law and the photon index is consistent with the
best-fit index of the power-law component from the \xmm\ fit (G09).
The low count statistics did not permit for a thorough verification
of the models using the \chisq-statistic, but a simultaneous fit of
the candidate qLMXB spectrum (\xmm\ and \chandra) showed that the
observed flux and spectral parameters of \sourceCeight\ and
\sourceCnine\ combined are consistent with those of the previously
observed \sourcefour.

The photon index of \sourceCnine\ is consistent with typical photon
indices of cataclysmic variables \citep[CVs;][]{richman96}.  A deeper
exposure will be required to attempt a more precise spectral fitting,
using thermal bremsstrahlung model for example, and confirm the
possible CV classification of this faint source.

Another candidate qLMXB, \sourcefive, was also observed in the field
of view and was named \sourceCseven.  The Cash-statistic fit of the
source spectra provided best-fit parameters that were consistent with
the values previously reported.  The unabsorbed flux (0.5-10 keV)
during the more recent \chandra\ observation, however, was a factor
$0.54\ud{0.30}{0.24}$ (90\% confidence) lower.  This is significantly
lower than the previously observed flux, and calls into question the
classification of this X-ray source as an H-atmosphere qLMXB, since
such strong variability is not expected on $\sim$years timescales,
unless a protracted ($\sim$years) long outburst (e.g., $L_{X}\approxgt$
\ee{37} \cgslum) ended recently (\approxlt 1 yr)
\citep{rutledge02c,brown09}; there is no evidence supporting this
scenario in the present case.

Finally, the improved \Chandra\ astrometry permitted us to verify the
association of the \xray\ source with its possible counterpart
\TwoMassfive\ with a probability of association of 99.66\%.

Overall, the \Chandra\ observatory allowed to resolve, both spatially
and spectrally, the single source in the core, \sourcefour, into two
sources, one of them being the candidate qLMXB.  The second observed
candidate qLMXB, \sourcefive, showed consistent best-fit H-atmosphere
parameters, but also exhibit a significant decrease in its flux, which
is not expected for qLMXBs.  A longer exposure will permit us to
assert with better certitude the spectral classification of the
sources in the core of NGC~6304 and the other candidate qLMXB.

\acknowledgements R.E.R. is supported by an NSERC Discovery grant. E.F.B.
and V.E.Z. acknowledge support from NASA under award no.~NNX06AH79G.  The
work of G.G.P. was supported by NASA grant NNX09AC84G. The authors also
thank the referee for his useful suggestions and for his prompt
replies.

\bibliographystyle{apj_8}
\bibliography{complete}

\clearpage
\newpage

\end{document}